\documentclass[conference]{IEEEtran}
\IEEEoverridecommandlockouts
\usepackage{cite}
\usepackage{amsmath,amssymb,amsfonts}
\usepackage{algorithmic}
\usepackage{graphicx}
\usepackage{textcomp}
\usepackage{xcolor}
\usepackage{hyperref}
\usepackage{multicol}
\usepackage{blindtext, subfig}
\usepackage{dblfloatfix} 
\def\BibTeX{{\rm B\kern-.05em{\sc i\kern-.025em b}\kern-.08em
    T\kern-.1667em\lower.7ex\hbox{E}\kern-.125emX}}

\begin{document}

\title{Data-driven Computing in Elasticity \\via Chebyshev Approximation
{}
}

\author{\IEEEauthorblockN{Rahul-Vigneswaran K\textsuperscript{*}\textsuperscript{\dag} \thanks{\textsuperscript{*}Website : \href{https://rahulvigneswaran.github.io}{rahulvigneswaran.github.io}} \thanks{ \textsuperscript{\dag}Work done while interning at CEN}}
\IEEEauthorblockA{\textit{Department of Mechanical Engineering} \\
\textit{Amrita Vishwa Vidyapeetham}\\
Amritapuri, India \\
rahulvigneswaran@gmail.com}
\and
\IEEEauthorblockN{Neethu Mohan, Soman KP}
\IEEEauthorblockA{\textit{Center for Computational Engineering and Networking (CEN)} \\
\textit{Amrita School of Engineering, Coimbatore}\\
Amrita Vishwa Vidyapeetham, India  \\
neethumohan.ndkm@gmail.com, kp\_soman@amrita.edu}

}

\maketitle

\begin{abstract}
This paper proposes a data-driven approach for computing elasticity by means of a non-parametric regression approach rather than an optimization approach. The Chebyshev approximation is utilized for tackling the material data-sets non-linearity of the elasticity. Also, additional efforts have been taken to compare the results with several other state-of-the-art methodologies.
\end{abstract}


\section{Introduction}
\label{introduction}
In the era where machine learning is catapulting the conventional methodologies into the future, computational mechanics is a field where there was not much of development until recently. Data-driven solvers have become an emerging field lately due to its obvious advantages over conventional heuristic solvers. Data-driven solvers attempt to utilize the material data-set, without empirically modelling a constitutive law based on the stress-strain relation of the material, upon which the conventional methodologies of computational mechanics were heavily dependent on. Conventional approaches get the job done but are not good enough because of the unpredictable mapping between the equations and the real world behaviours of the systems. One cannot convert every possible real-time scenario into an appropriate equation. This calls for the need of Data-driven solvers that function based on real-time data rather than a set of heuristically created equations. 

Elastic systems like shown in Fig. \ref{fig1} (a), are easy to solve due to its linear nature by using Hooke’s law \cite{1} while a system like shown in Fig. \ref{fig1} (b) are inherently difficult to solve due to its non-linearity of stress-strain relationship even at infinitesimally small strains inputs. There are several models used for solving nonlinear elasticity namely Ramberg-Osgood \cite{2}, Hyperbolic law \cite{3}, Hardin-Drnevich \cite{4}, Duncan-Chang \cite{5}, Duncan-Selig \cite{6} and the one that is commonly used is called Power law or also called as Ludwik’s law \cite{7}. Check \cite{19} for an extensive review. Even though the above-mentioned model capture the essence of nonlinear elasticity of the system to a certain degree, they are still a borderline approximation of the actual non-linearity. This is where date driven solvers like \cite{8,9,10,11,12,13,14,15,16,17,18} become superior over the conventional methods mentioned earlier. 

Section \ref{introduction} provides a basic introduction about the research work, Section \ref{related} explores and gives a detailed analysis of already existing literature and Section \ref{background} gives a background knowledge of topics such as nonlinear elasticity and Chebyshev polynomials for a better understanding of the work. Section \ref{experiments} provides an in-depth knowledge of the experimental setup and the methodologies utilized. Finally Section \ref{results} contains all the results obtained from the experiments and Section \ref{conclusion} gives a strong closing statement by pressing on the importance of this work, the significance of the results obtained and future scope

\section{Related Works}
\label{related}
Several methods like \cite{8,9} rely on minimizing the distance between a provided data set of the material and the subset of the appropriate stress-strain fields in equilibrium conditions. \cite{8} considers the Mahalanobis distance \cite{28} in order to incorporate the statistical uncertainty (2nd order) of data in computations. It is the distance between two groups' multiple means (centroids) used in the analysis of discriminant. In \cite{18}, the method of minimizing the distance is extrapolated for nonlinearity in elasticity and Lagrange multipliers enforce the physical constraints.
Data-driven solvers like \cite{10}, irrespective of the ideology of following a pure approach guided the given data in its entirety, heuristically adds a few constraints in order to develop a solver that is thermodynamically consistent during its solving phase. This directly contributes to fulfilling the second and first principles of thermodynamics which is independent of the experimental result’s quality. 
While most solvers lack generality due to its nature of the mathematical formulation, they fail at adapting themselves to new experimental data. This poses a real issue in real-time embedded applications as the solvers have to be remodelled periodically. \cite{11} overcomes this downfall by directly connecting the data to computers for performing simulations. The so-performed numerical simulations will employ rules based on universal laws while minimizing the need for specifically engineered models by making use of manifold learning \cite{29} methodologies. 

\section{Background}
\label{background}
\subsection{Nonlinear Elasticity}
\begin{figure}[htbp]
\centerline{\includegraphics[width=\linewidth]{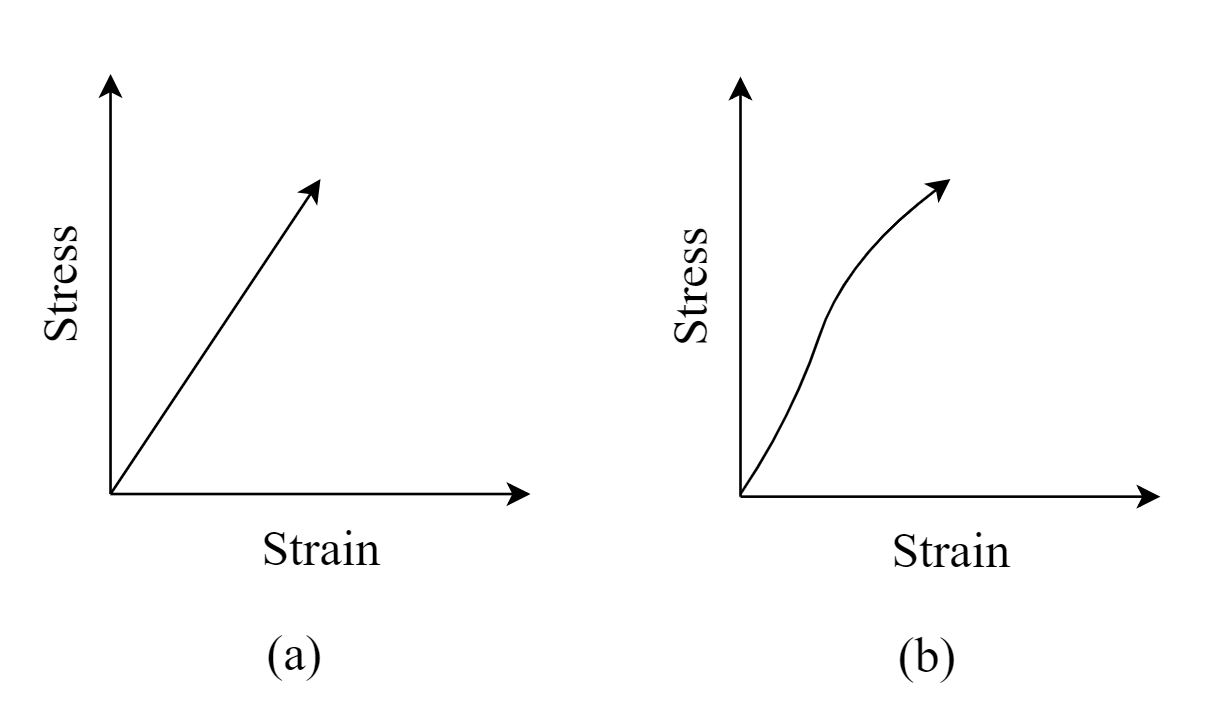}}
\caption{Stress-strain relation graphs of (a) Linearly Elastic and (b) Non-linearly elastic systems.}
\label{fig1}
\end{figure}

In order to understand nonlinear elasticity in detail, it becomes vital to have a basic understanding of elasticity in general. The key pillars for a material to be elastic in nature are – when the application of the load is stopped, the deformation vanishes, the deformation is independent of its history and the potential-energy exist as a function of deformation. When a material satisfies the above criteria for elasticity and its stress-strain relation is nonlinear like in Fig. \ref{fig1} (b), then they are called as nonlinear elastic materials.

This nonlinearity between the stress and strain is exactly the hurdle that prevents conventional methodologies from functioning at their fullest and makes it hard to map to a set of predefined equations.

\subsection{Chebyshev Polynomials}
Chebyshev polynomials play a vital role in approximation theory \cite{20} due to its roots of the Chebyshev polynomials of the first kind and are also known as Chebyshev nodes \cite{21}. During the polynomial interpolation, these nodes are used. As a result, the resulting polynomial of interpolation reduces the problem of Runge's phenomenon. It also delivers an approximation that is extremely near to the optimum approximation polynomial to a function that is continuous under the maximum norm. This approximation inherently leads to the Clenshaw–Curtis quadrature method \cite{22}. 

\begin{figure}[htbp]
\centerline{\includegraphics[width=\linewidth]{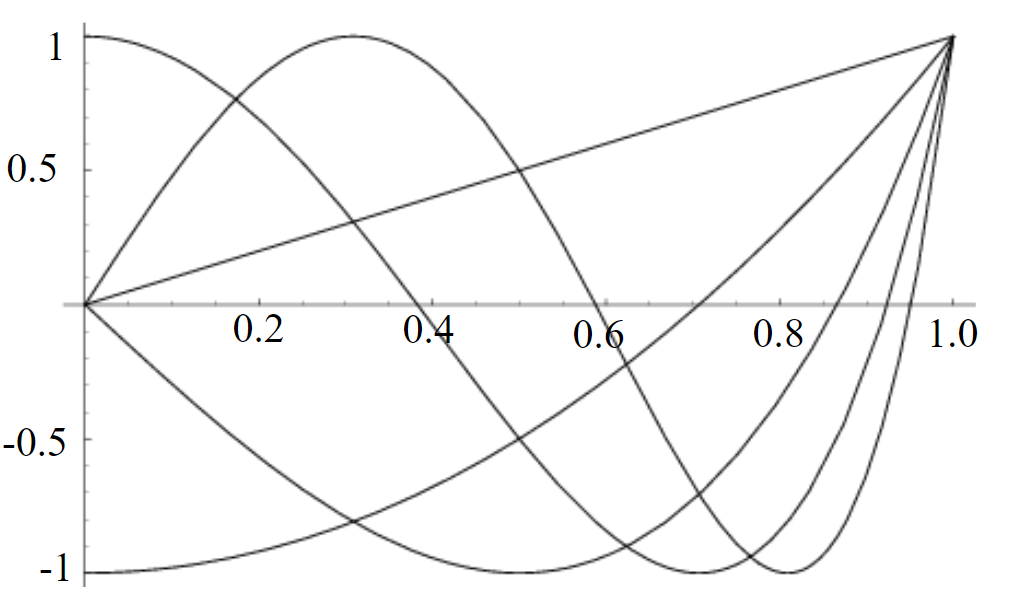}}
\caption{Illustration of The Chebyshev polynomials of the first kind ${T_n}\left( x \right)$ for $x \in \left[ {0,1} \right]$ and $n = 1,2, \ldots ,5$.}
\label{fig2}
\end{figure}

The Chebyshev polynomials of the first kind ${T_n}\left( x \right)$ are illustrated in Fig. \ref{fig2} for $x \in \left[ {0,1} \right]$ and $n = 1,2, \ldots ,5$. Using a Chebyshev Polynomial of the First Kind $T$, we can define 
\begin{align}
   {{c}_{j}} &\equiv \frac{2}{N}\sum\limits_{k=1}^{N}{f\left( {{x}_{k}} \right)}{{T}_{j}}\left( {{x}_{k}} \right) \\ 
  &=\frac{2}{N}\sum\limits_{k=1}^{N}{f\left[ \cos \left\{ \frac{\pi \left( k-\frac{1}{2} \right)}{N} \right\} \right]}\cos \left\{ \frac{\pi j\left( k-\frac{1}{2} \right)}{N} \right\}  
\tag{1}\label{tag1}
\end{align}

where, $k = 1,2, \ldots ,n$. Then, 

\[f\left( x \right)\approx \sum\limits_{k=0}^{N-1}{{{c}_{k}}{{T}_{k}}\left( x \right)}-\frac{1}{2}{{c}_{0}}\tag{2}\label{tag2}\]

For the $N$ zeros of ${T_N}\left( x \right)$ it is exact and this way of approximation is essential because, in case of truncation, the error is distributed evenly and smoothly over $\left[ { - 1,1} \right]$. The Chebyshev approximation \cite{23,30} is very near to the Minimax Polynomial. 

\section{Experiments}
\label{experiments}
\subsection{Experimental Setup}
We use MATLAB as the test platform and take advantage of chebfun \cite{27} package for carrying out calculations of Chebyshev approximation. For exponentially increasing the agility of processing of data, a computer with Intel Core i7-6700HQ Skylake CPU running at 2.60Hz and a DDR4 16 GB ram memory. The truss configuration used for the results shown in this paper is illustrated in Fig. \ref{fig3} but the methodology can be extended to any truss configuration. GitHub Repository for all the codes used in this paper:  \href{https://github.com/rahulvigneswaran/Data-Driven-Computing-in-Elasticity-via-Chebyshev-Polynomial}{https://github.com/rahulvigneswaran/Data-Driven-Computing-in-Elasticity-via-Chebyshev-Approximation}

\begin{figure*}[hb]
\centering
\subfloat[][]{\includegraphics[width=0.5\linewidth]{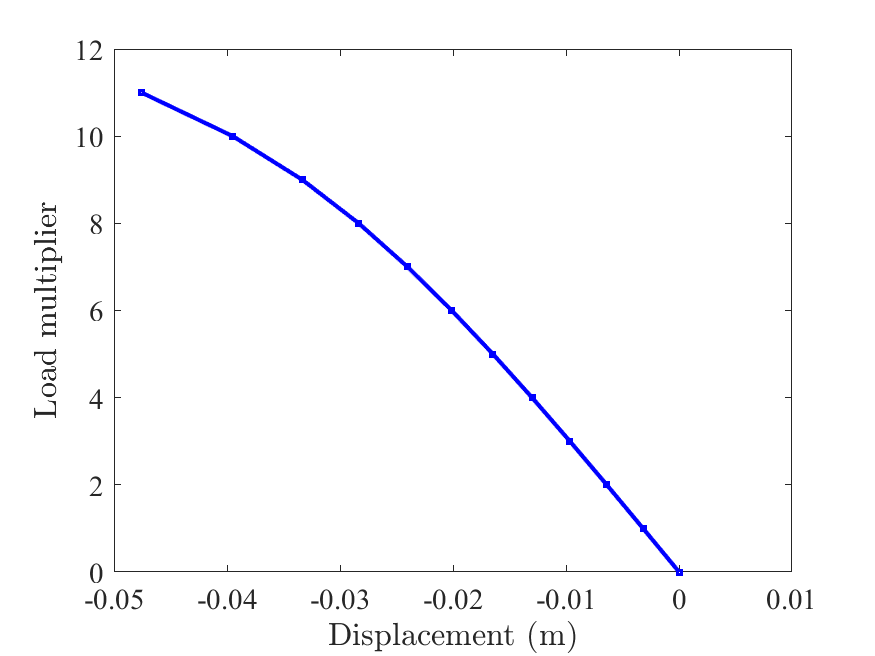}}
\subfloat[][]{\includegraphics[width=0.5\linewidth]{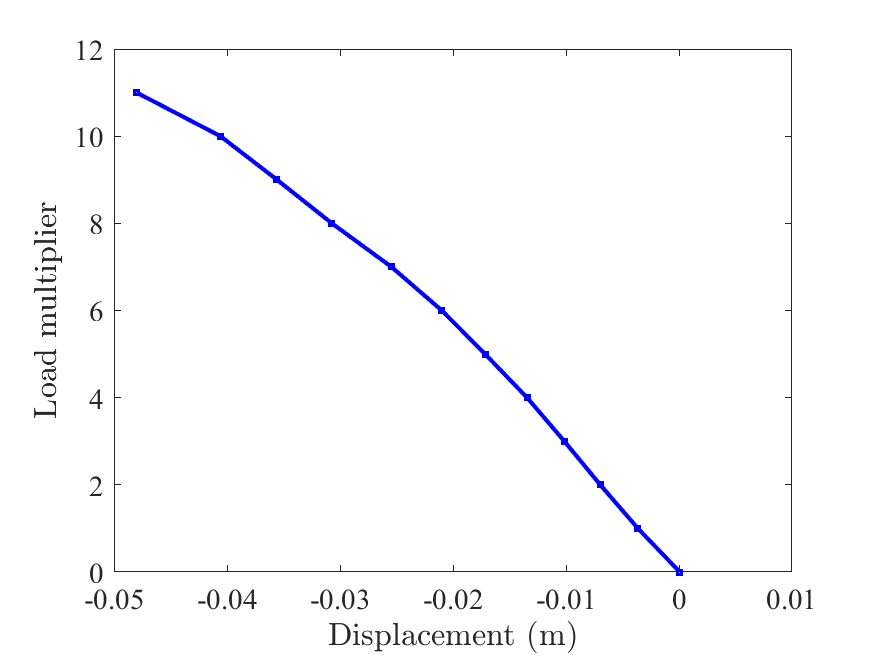}}
\caption{Obtained paths of equilibrium by varying load multiplier: (a) Method proposed in this paper; (b) Method proposed in \cite{26}.}
\label{fig5}
\end{figure*}

\begin{figure*}[hb]
\centering
\subfloat[][]{\includegraphics[width=0.5\linewidth]{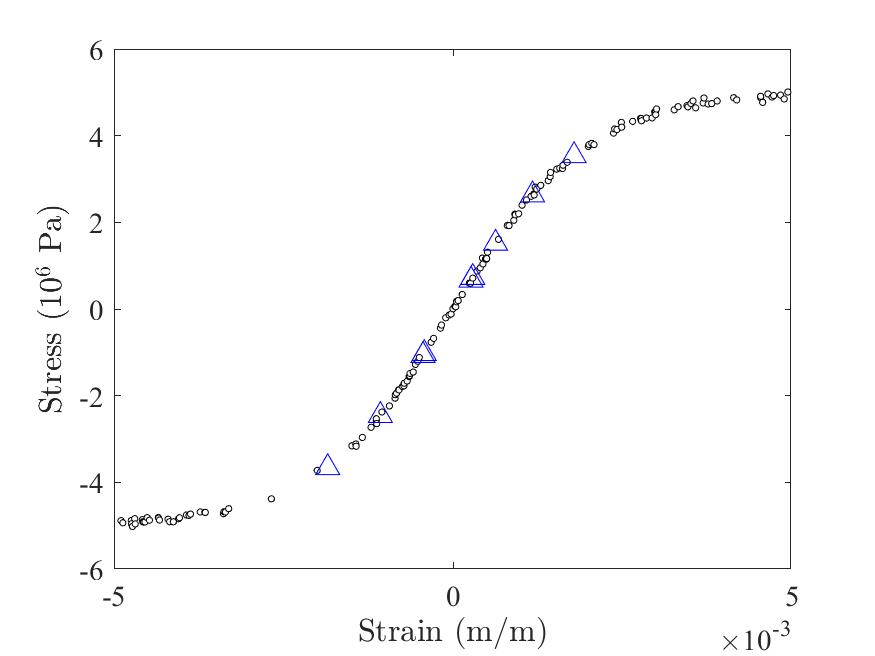}}
\subfloat[][]{\includegraphics[width=0.5\linewidth]{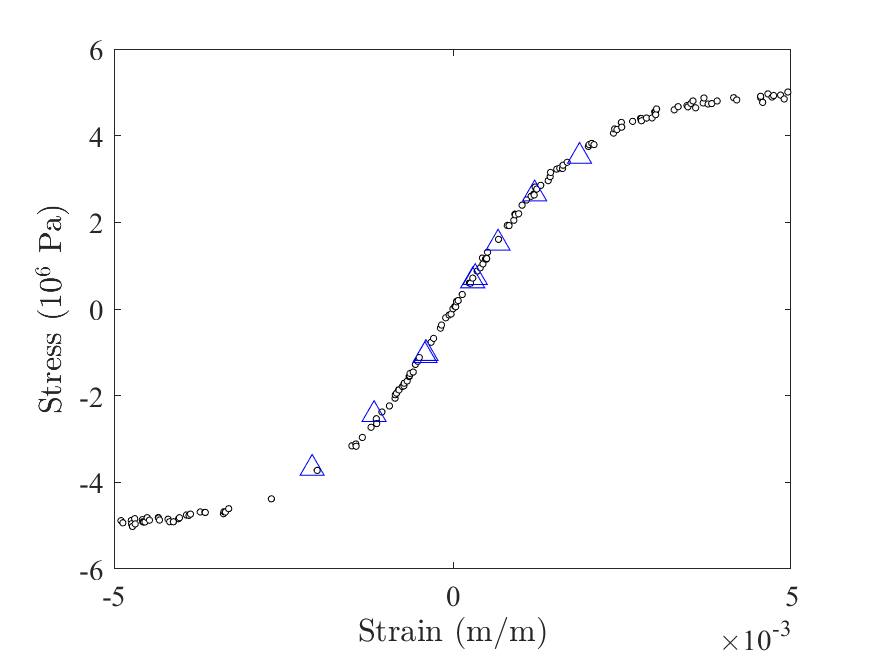}}
\caption{Solutions obtained for the load multiplier $\lambda = 10$: (a) Method proposed in this paper; (b) Method proposed in \cite{26}.}
\label{fig6}
\end{figure*}

\begin{figure*}[hb]
\centering
\subfloat[][]{\includegraphics[width=0.5\linewidth]{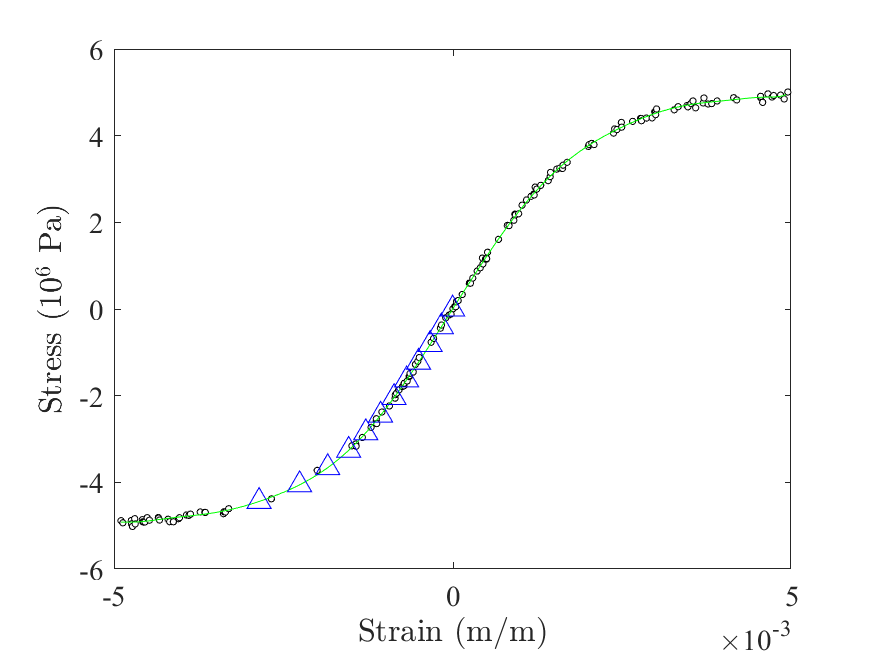}}
\subfloat[][]{\includegraphics[width=0.5\linewidth]{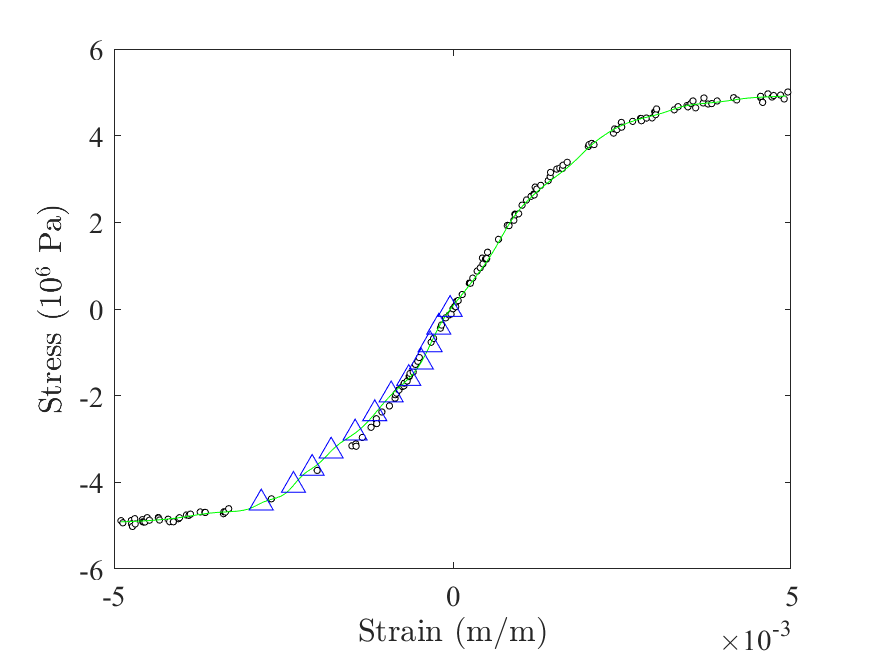}}
\caption{Obtained stress-strain pairs of member (A). (a) The proposed method and (b) Method from \cite{24} applied to the formulation in \cite{14}. $\bigcirc  \to $ material data point, $\Delta  \to $ obtained strain-stress pair.
}
\label{fig7}
\end{figure*}

\begin{figure}[htbp]
\centerline{\includegraphics[width=\linewidth]{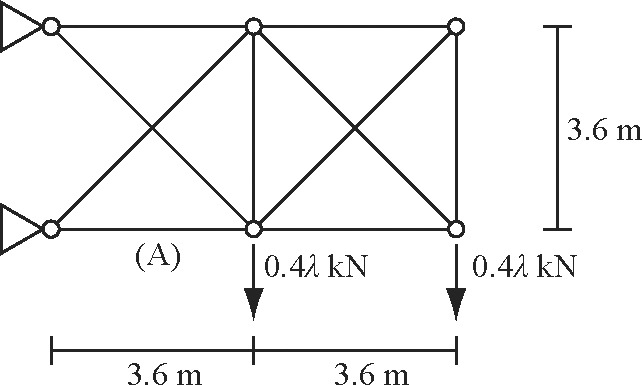}}
\caption{10-bar truss configuration \cite{26} }
\label{fig3}
\end{figure}

\subsection{Methodology}
Even though this methodology can be extended to any nonlinear elasticity problem unlike \cite{24}, for the purpose of simplicity, we have limited our explanation to the scenario of truss undergoing small deformations. This problem when formulated like in \cite{15}, inherently becomes a nonconvex optimization \cite{25} problem. In this paper, we have utilized the method used in \cite{26}, which proposes a novel method by converting the problem into a set of nonlinear equations that can be solved easily.

The compatibility conditions and the force balance relations are as follows,

\[\begin{gathered}
  {\varepsilon _i} = b_i^Tu \\ 
  \sum\limits_{i = 1}^m {{\upsilon _i}{\sigma _i}{b_i}}  = p \\ 
\end{gathered} \tag{3}\label{tag3}\]

where, ${\varepsilon _i}$ is axial strain, ${{\sigma _i}}$ is axial stress and $\upsilon _i$ is the volume of the member $i (i = 1,2, \ldots ,m)$ respectively. In equation \eqref{tag3}, $b _1,b _2, \ldots ,b _m$ are constant vectors. Lets suppose the material dataset has been provided and is represented as $D = \left\{ {\left( {{{\overset{\lower0.5em\hbox{$\smash{\scriptscriptstyle\smile}$}}{\varepsilon } }_1},{{\overset{\lower0.5em\hbox{$\smash{\scriptscriptstyle\smile}$}}{\sigma } }_1}} \right), \ldots ,\left( {{{\overset{\lower0.5em\hbox{$\smash{\scriptscriptstyle\smile}$}}{\varepsilon } }_d},{{\overset{\lower0.5em\hbox{$\smash{\scriptscriptstyle\smile}$}}{\sigma } }_d}} \right)} \right\}$ where, the observed uniaxial strains and stresses are denoted by ${{{\overset{\lower0.5em\hbox{$\smash{\scriptscriptstyle\smile}$}}{\varepsilon } }_j}}$ and ${{{\overset{\lower0.5em\hbox{$\smash{\scriptscriptstyle\smile}$}}{\sigma } }_j}}$ respectively. For each member $i$, the stress estimated $(S _i)$, for a given ${{\varepsilon _i}}$ is calculated by using the kernel regression \cite{31} in \cite{26} using the Gaussian kernel of the following form,

\[{s_i} = \frac{{\sum\limits_{j = 1}^d {\exp \left[ { - \alpha {{\left( {{\varepsilon _i} - {{\overset{\lower0.5em\hbox{$\smash{\scriptscriptstyle\smile}$}}{\varepsilon } }_j}} \right)}^2}} \right]} {{\overset{\lower0.5em\hbox{$\smash{\scriptscriptstyle\smile}$}}{\sigma } }_j}}}{{\sum\limits_{j = 1}^d {\exp \left[ { - \alpha {{\left( {{\varepsilon _i} - {{\overset{\lower0.5em\hbox{$\smash{\scriptscriptstyle\smile}$}}{\varepsilon } }_j}} \right)}^2}} \right]} }}\tag{4}\label{tag4}\]

where, $\alpha  > 0$ is acquired by cross validation of D. A data-driven solving approach can be formulated as,

\[minimize\left\| {\sigma  - s} \right\|\tag{5}\label{tag5}\]

with constraints \eqref{tag3} and \eqref{tag4}. Here, equation \eqref{tag4} in conjunction with the given material data, is basically used as a lookup table but the drawback is that the equation \eqref{tag4} has to be calculated every time the material data has to be referred by the algorithm. 

In this paper, we have experimented by replacing the equation \eqref{tag4}, that is, the Kernel regression with Chebyshev approximation (Equation \eqref{tag2}). The advantage is that, by doing so, we create a one-time smooth data curve from the given coarse experimental data which can be accessed easily like simple function $f(x)$ where $x$ is the input strain and $f(x)$ will provide the corresponding stress output. 

\section{Results}
\label{results}
\begin{figure}[htbp]
\centerline{\includegraphics[width=\linewidth]{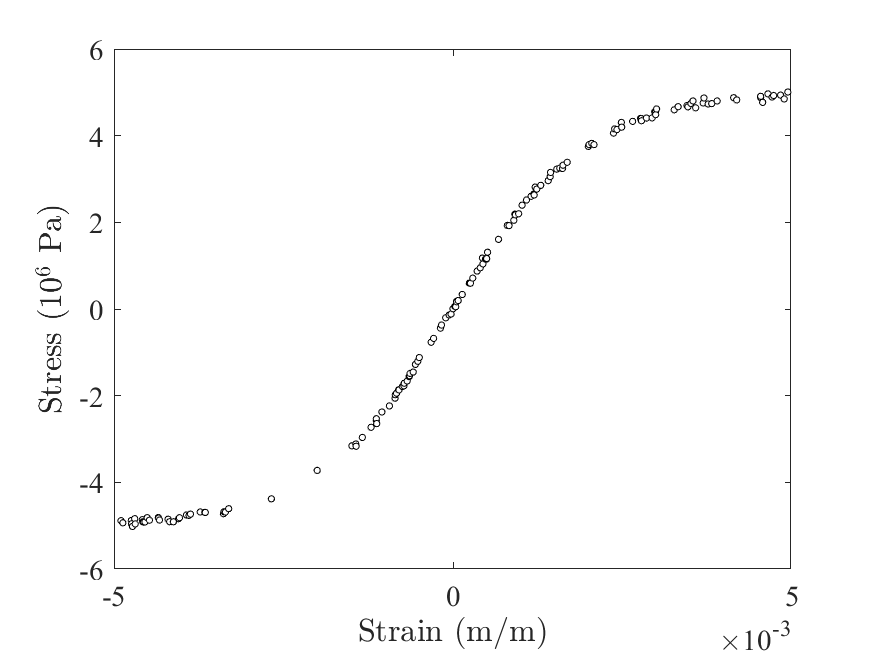}}
\caption{Material dataset used for the experiments.}
\label{fig4}
\end{figure}

An extensive comparison has been carried out for better assessment of the results. Fig. \ref{fig4} is the stress-strain relation of the material that is considered for the assessment. It is evident from it that the dataset is coarse in nature. The obtained equilibrium paths of the method proposed in \cite{26} and this paper are depicted in Fig. \ref{fig5}. The solutions obtained for the load multiplier $\lambda = 10$ are shown in Fig. \ref{fig6}. The obtained stress-strain pairs of member (A) in the truss (Fig. \ref{fig3}) are shown in Fig. \ref{fig7}.

\begin{table}[]
\centering
\caption{Computational Costs}
\begin{tabular}{|c|c|c|}
\hline
\textbf{Formulation} & \textbf{Solving Process} & \textbf{Time (s)} \\ \hline
Proposed Method & Chebyshev Approx. & 0.3 \\ \hline
 & Nonlinear Eqn. & 3.2 \\ \hline
 & \textbf{Total} & \textbf{3.5} \\ \hline
Kernal Regression \cite{26} & Cross Validation & 2.3 \\ \hline
 & Nonlinear Eqn. & 2.64 \\ \hline
 & Total & 7.18 \\ \hline
Formulation in  \cite{14} & MIP \cite{24} & 309.4 \\ \hline
Formulation in \cite{14} & Heuristic \cite{14} & 0.6 \\ \hline
Formulation in \cite{15} & NLP & 13.8 \\ \hline
\end{tabular}

\label{tab1}
\end{table}

Table \ref{tab1} compares the computation time of the proposed method with methods proposed in \cite{14,15,24,16}. It is evident from the result that the proposed method has significantly outperformed the previously proposed state-of-the-art model by $51.3\%$. Even though the model proposed in \cite{14} has a very low computation time compared to other models, it is heuristically designed. It does not necessarily yield an optimal solution. For example, for a load multiplier of 10, the value of the solution is 1.316 J, while the actual solution is 0.251 J. Thus, having 81\% deviation from the actual solution, proving that the heuristic converged an inexact solution, irrespective of that fact that it is computationally cheap. As a result, the method proposed in this paper has a significant advantage over the existing models, making it suitable for the application of nonlinear elasticity. A much more extensive analysis with varying load multipliers have been done and can be viewed at \href{https://rahulvigneswaran.github.io/Data-Driven-Computing-in-Elasticity-via-Chebyshev-Approximation/}{https://rahulvigneswaran.github.io/Data-Driven-Computing-in-Elasticity-via-Chebyshev-Approximation/}. 

\section{Conclusion}
\label{conclusion}
The same experiment was carried out several other algorithms including neural network based curve-fitting which showed significantly faster results compared to Chebyshev approximation. Yet, they were not considered as suitable candidates due to its inherent nature that requires it to fit a curve, which might become time consuming when the distribution of the data is sparse and the curve that is fit becomes increasingly erroneous and might not necessarily mimic the behaviour of the actual system while Chebyshev approximation can handle it with ease.

From the results, it becomes evident that the proposed methodology is significantly faster than the existing cutting edge models while preserving the generality approach in elasticity. As we can define the linear elasticity as a special subset of nonlinear elasticity, the proposed method holds good for both linear and nonlinear elastic cases, making it a universal solver for elastic bodies. Deep learning \cite{34} has shown its effectiveness in many of the emerging fields (\cite{32,33}) which were traditionally completely based on heuristic methods. Therefore, using Deep learning in conjunction with the methodology proposed in this paper can be considered as a future endeavour.

\section*{Acknowledgment}
The authors would like to show their acknowledgement to Dr Soman KP, head of CEN lab for exchanging his wisdom and providing with continuous support during the course of this research. Also, an immense gratitude is conveyed to the anonymous reviewers for their valuable insights.

\end{document}